\documentclass[aps,prl,preprint,groupedaddress]{revtex4-2}
\usepackage{amssymb}
\usepackage[pdftex]{graphicx}
\usepackage{epsfig}

\usepackage{caption}
\usepackage{subcaption}

\begin{document}

\title{Space-time coupling of the carrier-envelope phase in ultrafast optical pulses}

\author{Ido Attia, Eugene Frumker}
\email[]{efrumker@bgu.ac.il} 
\affiliation{Department of Physics, Ben-Gurion University of the Negev, Beer-Sheva 84105, Israel}



\begin{abstract}
The carrier-envelope phase (CEP)  plays an increasingly important role in precise frequency comb spectroscopy, all-optical atomic clocks, quantum science and technology, astronomy, space-borne-metrology, and strong-field science.
Here we introduce an approach for space-time calculation of the CEP in the spatially defined region of interest.
We find a significant variation of CEP in the focal volume of refracting focusing elements and accurately calculate its value.
We discuss the implications and importance of this finding.
 Our method is particularly suitable for application to complex, real-world, optical systems thereby making it especially useful to applications in research labs as well as in the engineering of innovative designs that rely on the CEP.
\end{abstract}

\maketitle
\section{Introduction}

Carrier-envelope phase (CEP), $\phi_{ce}$, which is defined as a phase offset between the rapidly oscillating electric field within ultrashort pulse and the peak of the pulse envelope \cite{Cundiff_femto_comb_RevModPhys_2003}, became ubiquitous in many branches of contemporary physics. For instance, in the frequency comb metrology \cite{Udem_OptFreqMetrology_Nature2002}, where the frequency comb offset ($f_0$) is linearly proportional to the pulse-to-pulse CEP shift \cite{Cundiff_femto_comb_RevModPhys_2003}, this CEP shift, along with pulse repetition rate ($f_r$) would define a precise frequency ruler - the frequency comb $f_n=f_0+nf_r$ \cite{Udem_OptFreqMetrology_Nature2002}.
 Such frequency combs, capable of spanning very broad spectral range all the way to the extreme ultraviolet \cite{Gohle_combs_XUV_Nature_2005}, already have had transformative impact on precise spectroscopy \cite{Schiller_Dual_comb_spectroscopy_OL_2002spectrometry, Coddington_multiheterodyne_spectr_combs_PRL_2008}, all-optical atomic clocks \cite{Udem_OptFreqMetrology_Nature2002}, quantum science and technology \cite{Roslund_ultrafast_comb_NatPhot_2014, Reimer_multiphoton_entangled_combs_Science_2016},  and even space-borne-metrology \cite{Lezius_space_combs_Optica_2016} and astronomy \cite{Steinmetz_combs_astronomy_Science_2008}.
CEP also plays  a central role in the strong field science \cite{Paulus_CEP_ionization_Nature_2001, Mashiko_DOG_HHG_CEP_2008}.
Hence, CEP metrology and control are of significant importance and a lot of effort has been dedicated to these goals \cite{Jones_CEP_modelocking_Science_2000, Kakehata_CEP_measure_single_shot2001, Baltuvska_CEP_OPA_PRL_2002, Wittmann_CEP_NatPhys_2009}.

Hitherto, in most of these works, albeit with few exceptions \cite{Tritschler_CEP_Gouy_pahse_OL_2005, Hoff_tracingCEP_NatPhys_2017, Zhang_observing_phaseVolume_PRL_2020}, it has been tacitly or explicitly assumed that CEP, $\phi_{ce}$, is strictly a temporal phenomenon, so that $E(t)=E_{en}(t)\cos(\phi_{ce}+2\pi f_c t)$, where $E_{en}(t)$ is the pulse  envelope and $\textrm{f}_c$ - denotes a carrier optical frequency.
In the early days of ultrafast science, the treatment of the interaction of femtosecond pulses with the matter was also considered as only a temporal phenomenon. However, it became evident that space-time coupling can't be neglected in the femtosecond \cite{Wefers_space_time_IEEE_1996, weiner_pulse_shaping_RSI_2000}
or attosecond science \cite{Frumker_sword_OL2009, Frumker_wavefronts_OE2012}.
Within focal volume the temporal profile of the pulse may vary significantly \cite{Walmsley_charact_em_pulses_AdvancesOptPhot_2009, Selcuk_SpaceTime_J_Opt_2010, Frumker_wavefronts_OE2012}. Taking into account space-time coupling is critical for analysing experimental data, design, control and avoiding artifacts in the ultrafast science experiments \cite{Herek_quantum_flow_space_time_Nature_2002, Walmsley_charact_em_pulses_AdvancesOptPhot_2009}.

Recently, an analytical approach has been introduced to estimate the CEP distribution in the focal volume \cite{Porras_charact_E_field_OL_2009, Porras_3D_CEP_PRA_2018}. While providing valuable insight, these analytical tools are limited only to very simple optical systems and field distributions.

In this work, we demonstrate a new numerical method for space-time calculation of the CEP, $\phi_{ce}$. Our approach is not limited to simple apparatus, rather is well suited for applications to complex, real world, optical systems.  We consider several basic key examples of focusing optics, calculate explicitly CEP within the region of interest and discuss implications of our findings.

\section{Method}
Evolution of ultrashort pulse electric field in space and time, $\vec{E}(\vec{r},t)$, can be decomposed into its spectral components, $\vec{\tilde{E}}(\vec{r},f)$, by a Fourier transform:

\begin{equation}
 \label{Eq_FT}
   \vec{\tilde{E}}(\vec{r},f)=\int_{-\infty}^{\infty} \vec{E}(\vec{r},t)e^{-i2\pi f t} dt \equiv \mathcal{F}\{\vec{E}(\vec{r},t)\}.
\end{equation}

 For the sake of simplicity, we will assume scalar approximation ({i.e. assuming that $\vec{E}(\vec{r},t) = E(\vec{r},t) \hat{x}$) , even though our approach doesn't necessitate this and can be used for analysis of fully vectorial fields. We assume ultrashort pulse propagation along the $\hat{z}$ axis.
 Ultrashort laser pulse, $E(x,y,z=0,t)$, entering an optical system at $z=0$ is decomposed into monochromatic fields, $\tilde{E}(x, y, 0, f)$, using Eq.\ref{Eq_FT} at the entrance of the system to be analysed.
 Each of these monochromatic field components of the pulse is propagated using optical code for the propagation of the continuous wave (CW) light fields into spatial region of interest $\vec{r} \in \mathbb{R}_{\textrm{interest}}^3 $. For example, for many scientific and technological applications, such a region of interest is a focal volume of an optical system.
 Then, for each point $\vec{r}$ in the region of interested, the complex frequency field components are recomposed together using inverse Fourier transform, $ E(\vec{r},t)=\int_{-\infty}^{\infty} \tilde{E}(\vec{r},f)e^{i2\pi f t} df \equiv \mathcal{F}^{-1}\{\tilde{E}(\vec{r},f)\}$, to obtain a temporal laser field evolution of the optical pulse.

 In our work, we have chosen to use commercially available optical codes, intended primarily for lens design, for the complex light field propagation.  Using such widely used commercial optical codes has several distinct advantages. First,  these codes are usually suitable for analysis of real world optical systems of significant complexity and are not limited to simple systems. For example, field propagation in optical systems consisting of many lenses, mirrors, prisms, gratings, etc. can be easily performed. Second, there are several optical codes available, which have been well tested, debugged and optimized during many years of real-world scientific and industrial applications. Thus we leverage these advantages in our approach.
  However, these optical codes, capable of physical optics  simulations (i.e propagation of complex optical field and not only ray optics) are usually intended for use for the propagation of the continuous wave (CW) rather than ultrashort optical pulses. Decomposing ultrashort pulse into its monochromatic spectral components allows to easily adopt the CW optical codes for the analysis of ultrashort pulses \cite{Fuchs_propagation_OptExp_2005}. Specifically, we used OSLO \cite{Shannon_Opt_Design_OSLO_book1997} code in this work, but any other code for CW optical field propagation could be used instead, for example Zemax \cite{Geary_Intro_Zemax_book_2002} or CodeV \cite{codeVoptical}.

Once the temporal evolution of the pulse is reconstructed in the spatial region of interest, we find the CEP, $\phi_{ce} (\vec{r})$,  by extracting the phase between the peak of the pulse envelope, at time $t_{\textrm{max}}$, and the closest peak of the carrier wave, by a conventional formula $\phi_{ce} (\vec{r}) = \textrm{Arg}(E(\vec{r},t_{\textrm{max}}))=\arctan(\textrm{Re}(E(\vec{r},t_{\textrm{max}})), \textrm{Im}(E(\vec{r},t_{\textrm{max}})))$.
 The result is unwrapped - whenever the reconstructed CEP jumps between adjacent points in space more than or equal to $\pi$ radians, the multiples of $2\pi$ is added until the jump is less than $\pi$.


  To facilitate numerical calculations, instead of the continuous Fourier transform (Eq. \ref{Eq_FT}), we use a Discrete Fourier Transform (DFT). Working with DFT has several significant consequences that should be taken into account in the context of our analysis to achieve efficient calculations and to avoid distortion and aliasing artifacts.
  The spectral amplitude $\tilde{E}(\vec{r},f)$ of the pulse is sampled within the pulse bandwidth at $N$ points with frequencies $f_{sample}=[f_{min},f_{min}+ \Delta f,..., f_{min}+ (N-1) \Delta f ]$.
  The sampling bandwidth has to be chosen to represent the spectral bandwidth of the actual pulse accurately. Choosing the bandwidth too narrow will result in pulse distortion in the time domain, and choosing the bandwidth that is much wider than the actual pulse bandwidth would result in inefficient use of the computer resources.

  Special care should be taken in choosing the sampling spacing  $\Delta f$ to minimize the aliasing \cite{Brigham_FFT_1988} effect.
  Using a discrete sampling instead of continuous spectral amplitude, $\tilde{E}(\vec{r},f) \rightarrow \tilde{E}(\vec{r},f) \times \sum_{n=0}^{N-1} \delta (f-[f_{min}+n \Delta f])$, results in periodic pulses in time domain with a period $ T=1/\Delta f$. As the pulse propagates along optical system, it often disperses to some characteristic time scale $\tau_{max}$. The $\tau_{max}$ depends both on material dispersion and on the system geometry. In the case of Fourier limited pulse, the shorter the pulse at the entrance of the system, the wider bandwidth it has and more dispersion it will experience that will result in the larger $\tau_{max}$.   To avoid the aliasing, it is important to choose $T>\tau_{max}$, implying that the sampling spacing should be chosen so that $\Delta f < 1/\tau_{max}$.
  Strictly speaking, finite pulse bandwidth necessarily implies infinite pulse duration in the time domain. However, the pulse will decay in time to negligible amplitude. So intelligent judgment has to be made to set the frequency sampling spacing dense enough so that aliasing will become negligibly small and at the same time not to over-sample to avoid wasting the computer resources.

  Note that frequency comb spacing ($\Delta f$) used for calculations should not necessarily be identical to the frequency spacing ($ f_r$) of the real laser system frequency comb. For example, the repetition rate of commonly used CEP stabilized femtosecond amplifiers is just of the order of 1 kHz, implying
  $ f_r = 1$ kHz. Such a dense sampling of typical femtosecond pulse spectra would be impractical due to the computer's memory limitations and is not needed to achieve accurate space-time pulse evolution computations and the CEP retrieval.

The frequency spectrum of a typical femtosecond pulse usually spans  near-infrared, visible and sometimes ultraviolet (UV)  regions.
Hence, there is no spectral content in the pulse from zero to about hundreds of Terahertz. To make computations more efficient, it is useful to eliminate this zero-padded region of spectra by  shifting
the pulse spectral envelope, $\tilde{E}_{env}(\vec{r},f)$ to zero, noting that $\tilde{E}(\vec{r},f)=\tilde{E}_{env}(\vec{r},f) \ast \delta (f-f_c)$. After space propagation of the $\tilde{E}_{env}(\vec{r},f)$, applying the convolution theorem we reconstruct the pulse in the time domain, $E(\vec{r},t)=E_{env}(\vec{r},t) \times \exp{ \{ i 2\pi f_c t\}}$.

  The DFT is implemented in the software using a Fast-Fourier Transform (FFT) algorithm \cite{Brigham_FFT_1988}. The FFT algorithms work much more efficiently when the length of the vector to be transformed is a power of 2 of a positive integer number ($N=2^m, m \in \mathbb{N}$). We achieve this condition by slightly reducing spacing in the frequency domain or applying zero-padding in the time domain.

 \section{Results}

  We start by analyzing several simple well-known scenarios in order to confirm the validity of our approach and of the computer simulations.
%
%

   Good example of such a scenario is focusing of the femtosecond pulse by  a lens as shown in the Fig. \ref{fig:aspheric y-z}. In this example, the lens is an aspheric  Geltech lens made of C0550 Corning glass with effective focal lens $\textrm{f}_{\textrm{EFL}}=3.1$ mm. The aspheric surface is chosen to correct spherical aberration, so that chromatic aberration is the main remaining aberration  \cite{Fuchs_propagation_OptExp_2005}. Further details about this aspheric lens are given in the supplemental document. The 25 fs pulse with Gaussian spatial profile impinges on the lens so that the waist of each spectral component of the pulse has the same 2.5 mm width and is situated just in front of the lens.

Figure \ref{fig:aspheric y-z} (a) shows a spatial field distribution of the pulse, 440 fs before the focusing. Focusing is defined as an instance in time with maximum pulse intensity. An emblematic "horseshoe" shape and a forerunner pulse are observed as expected \cite{Bor_Distortion_Femto_OptCom_1992}. Figure \ref{fig:aspheric y-z} (b) shows the same pulse 235 fs after the focusing. Also, a typical "v" shape of the femtosecond pulse spatial field  distribution after the focus \cite{Bor_Distortion_Femto_OptCom_1992} was clearly observed  as well.
\begin{figure}[h!]
  \includegraphics[width=\linewidth]{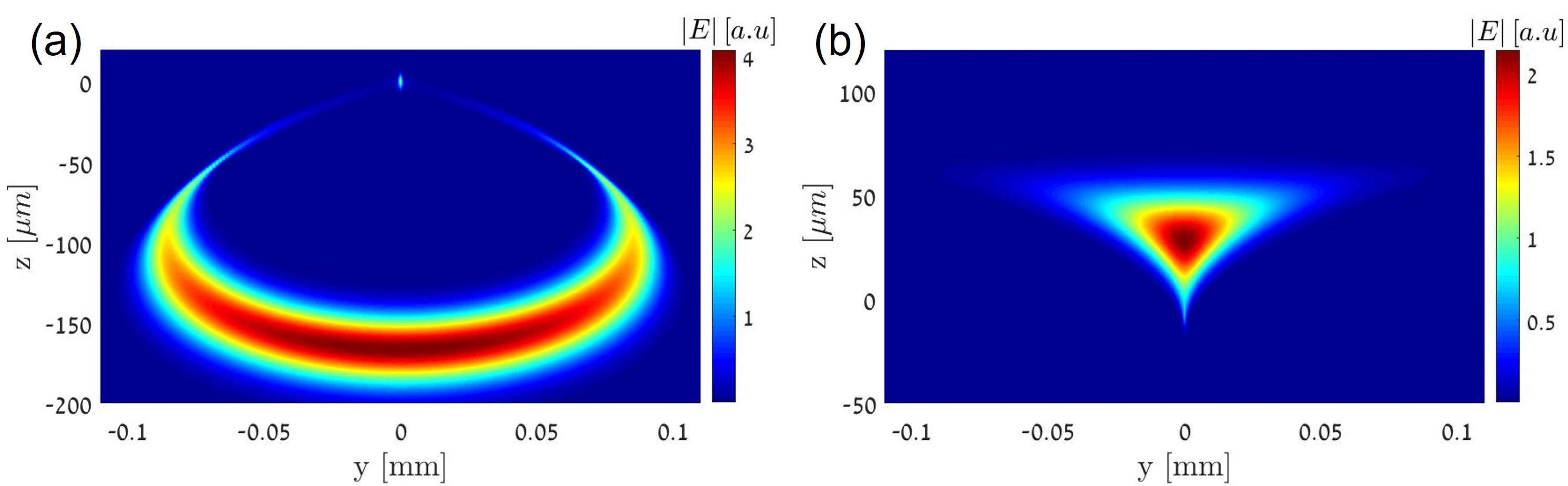}
  \caption{ Femtosecond pulse behaviour in the focal region focused by aspheric Geltech lens: (a) 440 fs before focusing; (b) 235 fs after the focusing.}
  \label{fig:aspheric y-z}
\end{figure}
%
%
%

Now we turn our attention to the CEP numerical experiments with different optical systems.  We study the following optical systems: (1) An aberration-free  "perfect" lens that has only "ideal" focusing effect on incoming beam regardless of numerical aperture and wavelength; (2) Optical 5 mm thick slab made of C0550 glass; (3) Spherical focusing mirror. Such a mirror induced geometrical aberrations but has no chromatic aberration nor dispersion; (4) Refractive focusing element - a lens. Here we considered the same aspheric Geltech lens as defined in the previous section.


\begin{figure}[h!]
  \includegraphics[width=\linewidth]{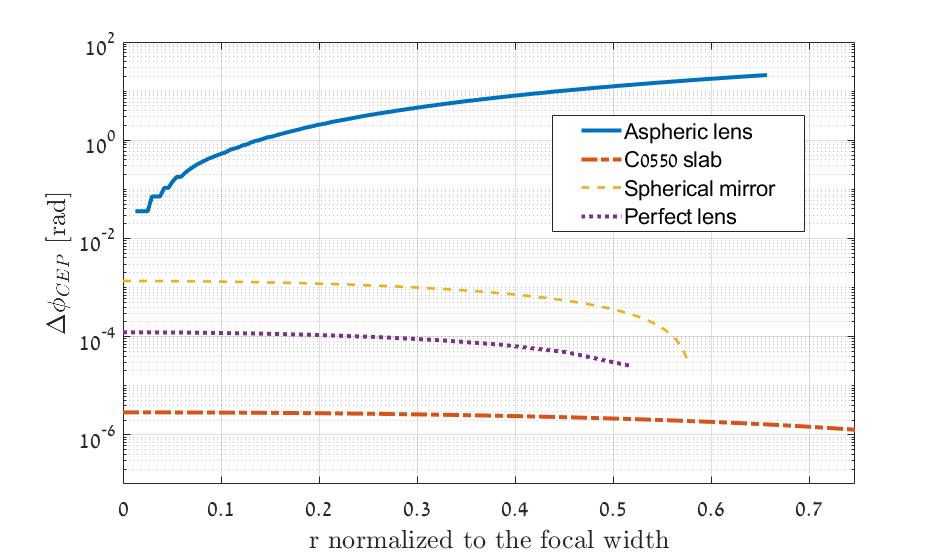}
  \caption{ Log scale of the CEP variation $\Delta \phi_{CEP}$ as the function of the radial distance in the focal plane for different optical systems: Geltech aspheric lens, C0550 slab, spherical mirror, and perfect lens. }
  \label{fig:CEP}
\end{figure}

Figure \ref{fig:CEP} shows the CEP variation across the focal plane for these systems.
 For the perfect lens (dotted line) and  spherical focusing mirror (dashed line), the CEP variation is negligible (within computational numerical error) regardless of the "tightness" (NA) of the focusing. Propagation of collimated femtosecond pulse through dispersive glass (dash-dotted line) shows a similar result - no variation of CEP across the beam.
It is an expected result, while propagations via dispersive material result in CEP shift due to difference between phase and groups velocities, in collimated beam this CEP shift will be largely the same across the beam, hence no meaningful spatial variation of the CEP is expected. The small observed CEP variation ($<10^{-3}$ rad) results from subtle diffraction of the collimated beam propagating through the slab.
In the case of a spherical focusing mirror, the CEP variation is larger than in previous cases but still negligible ($\sim10^{-3}$ rad) for any practical purpose.
However, the situation changes entirely for the refracting focusing element. A Geltech aspheric lens (solid line in the Fig. \ref{fig:CEP}) shows significant CEP variation across the focused beam.

\begin{figure}[h!]
  \includegraphics[width=\linewidth]{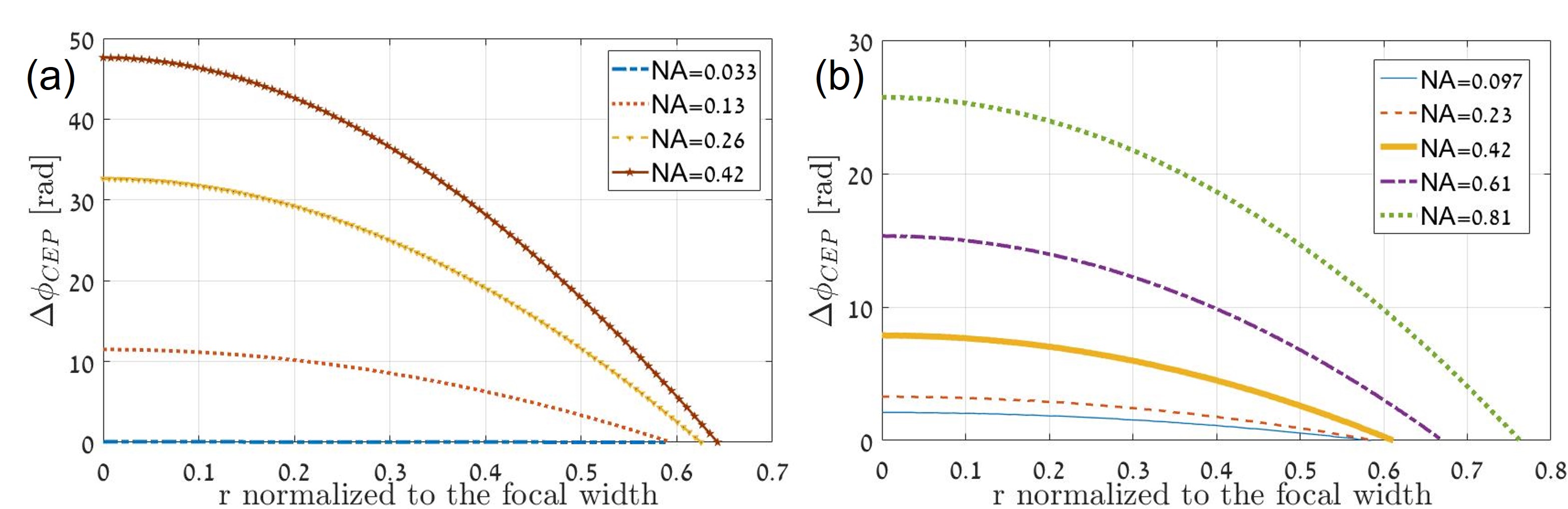}
  \caption{CEP variation, $\Delta \phi_{CEP}$, as the function of the radial distance in the focal plane normalized by the spot size at the focus is shown for: (a)  The achromatic doublet; (b) 
   The Geltech aspheric lens.
    Results are shown for different focusing conditions.}
  \label{fig:CEP(NA)-united}
\end{figure}
%
%
%

As the refractive aspheric lens has exhibited significant CEP variation across the focal plane, we add to our analysis another   refractive focusing element often used in the femtosecond optics - an achromatic doublet (for example, in the refractive femtosecond pulse shapers \cite{weiner_pulse_shaping_RSI_2000}). In this example, we have chosen an uncoated AC254-030 lens by Thorlabs with a focal length $\textrm{f}=30$ mm. Further details about this achromatic doublet lens are given in the supplemental document.
While for an achromatic doublet the chromatic aberration is notably compensated, a femtosecond pulse propagation through such a lens experiences significant material dispersion.
 We explore how CEP spatial variations within the focal volume of a lens depend on the focusing conditions. Figure \ref{fig:CEP(NA)-united} shows CEP variation for the aspheric lens (panel (a)) and for the achromatic doublet (panel (b)) for different NA of the focusing system. For very "relaxed" focusing conditions, such as $NA \simeq 0.03$, which corresponds to the waist size of w=1 mm for the achromatic doublet, there is almost no variation of the CEP phase within the focal volume. However, when the focusing becomes tighter, the variation of CEP within the focal volume increases significantly.

Even at still very gentle focusing conditions of NA=0.13, the variation of the CEP within the focal volume exceeds $2 \pi$ radians for the achromatic doublet. These variations of CEP are very significant; to put it into perspective, the difference between the cosine-like pulse and sine-like pulses is only  $\pi/2$ radians.
Note that even at small NA values ( example of $\textrm{NA} \simeq 0.1$ is shown in the Fig. \ref{fig:CEP(NA)-united} (b)) for the short focal length aspheric, meaning the impinging beam width of only few hundred microns), there is still meaningful residual variation of the CEP, particularly near the periphery of the focal volume. Further reduction of the impinging beam width doesn't translate into vanishing CEP phase as occurs for the long focal length lens (Fig. \ref{fig:CEP(NA)-united} (a)). We attribute this to the diffraction effect that becomes significant as the beam width approaches the wavelength scale of the incoming laser pulse.

Our results indicate that it is really an interplay between geometrical (focusing) effect,  diffraction and chromatic dispersion that leads to CEP variation across the beam- any of these contributions alone would not cause spatial CEP variations.

\begin{figure}[h!]
  \includegraphics[width=\linewidth]{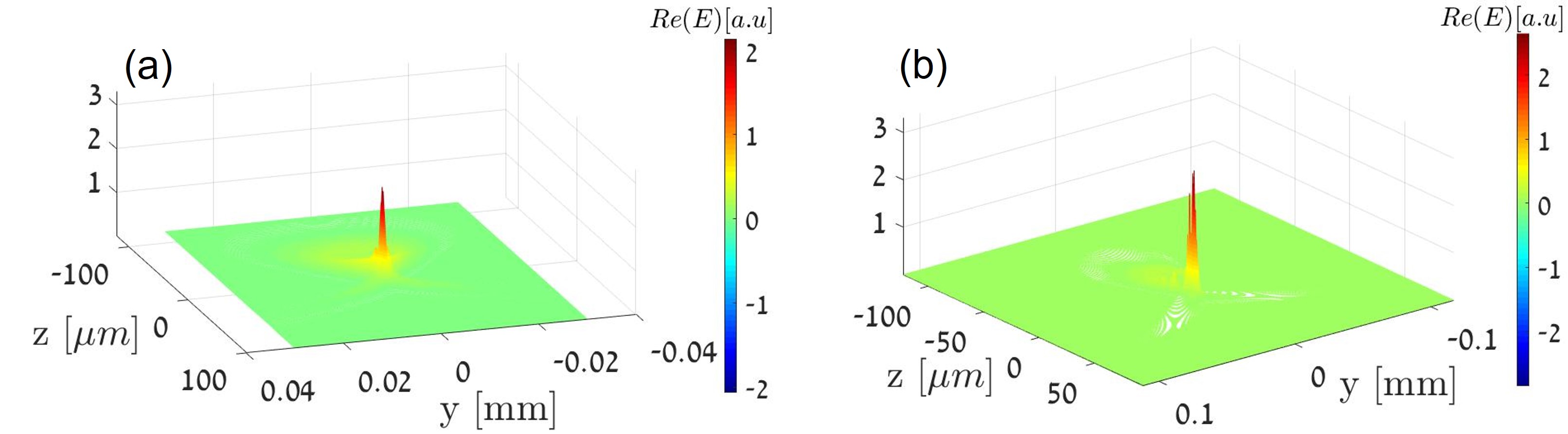}
  \caption{Spatial distribution of the field in the focus of: (a) Achromatic doublet, NA=0.42; (b) Geltech aspheric lens, NA=0.8.  Example of the space-time evolution "movie" of the field at the focus appears in the supplemental document Visualization 1.}
  \label{fig:CEP field}
\end{figure}

%

Spatial distributions of the electrical field at the focus for the aspheric lens and achromatic doublet are shown in the Figure \ref{fig:CEP field} for their highest respective NAs. Note that a significant field curvature in both cases accompanies substantial variations in the CEP within a focal volume. The characteristic horseshoe shape \cite{Bor_Distortion_Femto_OptCom_1992} in the focus is clearly visible for both lenses.
Similar to the increase of CEP variation with an increase in NA, we observe an increase in the field curvature as well as in the horseshoe shape appearance as we increase the focusing NA for both lenses.

\section{Discussions and conclusions}

Our results show that there is a significant variation of CEP in the focal volume of the refracting focusing element.
Spatial CEP variations strongly increase with a numerical aperture. We have observed a strong correlation between the CEP spatial variations and curvature of the pulse wavefront within the focal region.  Use of refractive optics is quite common and ubiquitous in ultrafast science and applications, particularly with lower energy pulses.
As we do not observe spatial CEP  variations with a focusing mirror, where just a reflection plays a role, our results indicate that it is a combination of geometrical focusing and dispersion that causes such a phenomenon.
Our findings imply that if one observes or drives physical phenomena dependent on the CEP with such an optical system, the result may be obscured or washed out by averaging within a focal volume. This means that it is essential to estimate the spatial behaviour of CEP within a region of interest.

To address this challenge, we have introduced an approach for a robust and accurate quantitative evaluation of the spatial behaviour of CEP and the pulse field evolution in space and time. Our method is not limited to simple setups. It can particularly useful as a powerful tool for designing and engineering of the ultrafast optical systems, where CEP matters.

In future work, it will be very interesting and useful to synergistically use both analytical methods \cite{Porras_charact_E_field_OL_2009, Porras_3D_CEP_PRA_2018} and our numerical approach. Measuring the spatial distribution of CEP in the lab is a non-trivial \cite{Tritschler_CEP_Gouy_pahse_OL_2005, Hoff_tracingCEP_NatPhys_2017}, time and resource-consuming task. Instead, our method can be used, providing the results much faster at a fraction of cost and effort, when compared to the actual lab experiments. Thus, our approach will drive major advances not only in the CEP related experimental work, but also in the development of the theory and the development of new techniques for suppression of the CEP variations in the focal volume.

\section{Acknowledgements}

This work has been supported by Israel Science Foundation (grant No. 2855/21) and European Commission Marie Curie Career Integration Grant.

\section{Disclosures}
The authors declare no conflicts of interest.

%

See Supplement 1 for supporting content.
%

\bibliography{G:/NRC_work_since_December2007/PostDoc_Papers/Orientation_nature/Orientation_Science/Atto_references}
\end{document}